\documentclass[10pt]{llncs}
\usepackage{psfig}
\def\be{\begin{equation}}
\def\ee{\end{equation}}

\begin{document}

\title{What Information Theory Can Tell Us About Quantum Reality}
\author{C. Adami and N.J. Cerf}
\institute{W. K. Kellogg Radiation Laboratory\\
         California Institute of Technology,
Pasadena, California 91125, USA}


\maketitle

\begin{abstract}
   An investigation of Einstein's ``physical'' reality and the concept
  of quantum reality in terms of information theory suggests a
  solution to quantum paradoxes such as the Einstein-Podolsky-Rosen
  (EPR) and the Schr\"odinger-cat paradoxes. Quantum reality, the
  picture based on unitarily evolving wavefunctions, is complete, but
  appears incomplete from the observer's point of view for fundamental
  reasons arising from the quantum information theory of measurement.
  Physical reality, the picture based on classically accessible
  observables is, in the worst case of EPR experiments, unrelated to
  the quantum reality it purports to reflect. Thus, quantum
  information theory implies that only correlations, not the
  correlata, are physically accessible: the mantra of the Ithaca
  interpretation of quantum mechanics.

\end{abstract}

\section{Introduction}
\noindent The concept of ``physical reality'' as championed by
Einstein~\cite{bib_epr}---the postulate that the {\em objective} state
of a system is specified by a set of real-valued parameters {\em
  independently} of our knowledge of them---has been an object of
contention ever since the inception of quantum theory (see, e.g.,
\cite
{bib_bohrnat,bib_wheeler,bib_bellbook,bib_schommers,bib_mittel,bib_cushing}).
The most prevailing views assert either that the ``quantum reality''
suggested by wavefunctions and non-local correlations is only a
mathematical construction necessary for a consistent theory (Bohr's
view), or else that physical reality is deterministic but incompletely
described by quantum mechanics (Einstein's view).  A popular
interpretation of the latter view is that physical reality is obscured
by inaccessible hidden variables~\cite{bib_bohm2}, a stance that
appears to be discredited by the violation of Bell's inequalities in
quantum mechanics~\cite{bib_bell}. Bohr's view of complementarity, on
the other hand, assigns a special status to classical physics as an
essential ingredient in measurement since it requires the measurement
device to be classical. As recognized by von Neumann~\cite{bib_vn},
this undermines the foundations of quantum mechanics as a complete and
consistent theory. Here, we suggest that Einstein realism and Bohr's
complementarity principle can be reconciled within a framework that
consistently describes the concept of information in quantum
mechanics. This is exemplified by the quantum information theoretic
treatment of the Einstein-Podolsky-Rosen (EPR)
experiment~\cite{bib_epr} and the Schr\"odinger-cat
paradox~\cite{bib_schroed}, which has recently attracted increasing
attention (see, e.g., \cite{bib_monroe}).  We propose that, in
general, the perceived physical reality and quantum reality can be
{\em disjoint}, that is, the result of a quantum measurement
conceivably might not carry any information---in the sense of Shannon
theory~\cite{bib_shannon}---which would allow the observer to infer
the state of the measured system.  While counterintuitive, we shall
show that this picture is a direct consequence of an
information-theoretic reinterpretation of quantum measurement.
Moreover, such a view effortlessly resolves the EPR paradox which has
inspired the discussions on reality, as well as other quantum
paradoxes rooted in the measurement problem.

The gedankenexperiment that constitutes the EPR paradox was created by
Einstein, Podolsky, and Rosen to demonstrate their dissatisfaction
with ``unknowables''~\cite{bib_epr}. In that experiment, it appears that
two complementary variables (such as position and momentum) are {\em in
principle} measurable by exploiting the quantum correlations between
the two particles, in contradiction with Heisenberg's uncertainty
principle.  Their conclusion, namely that the quantum mechanical description
of reality must therefore be incomplete, was based on a criterion for reality
which they considered ``reasonable'' (see below).  This criterion
was faulted by Bohr~\cite{bib_bohr} in his reply to the EPR paper,
insisting rather that physical variables are never independent of the
way they are measured owing to the complementarity principle, and
therefore that measurements do not confer reality to properties of
quantum objects.  We shall show here, using quantum information
theory 
only, that, while indeed an element of
reality is {\em not} created for the measured quantum system, the result of a
quantum measurement creates an element of reality for the result
of {\em another} measurement, i.e., it allows you to predict the
state of another measurement {\em device} without revealing the state
of the quantum system itself. Thus, physical reality is reflected in {\em
  correlations} between classical objects only. This view, which we
arrived at from a quantum information-theoretic examination of quantum
measurement~\cite{bib_ca2,bib_meas} essentially coincides with
Mermin's ``Ithaca Interpretation of Quantum Mechanics'',
Ref.~\cite{bib_mermin}.

\section{The EPR Paradox}
The EPR experiment in the version of Bohm~\cite{bib_bohm}
involves the preparation of a quantum system such as the one created
by the decay of a spinless particle into two half-integral-spin
particles: 
\be 
|\Psi_{\rm EPR}\rangle =
\frac1{\sqrt2}\left(|\uparrow\,\downarrow\,\rangle -
|\downarrow\,\uparrow\,\rangle\right)\;.\label{epr} 
\ee 
This state
represents the {\em superposition} of the two possible situations:
``left-particle spin-up, right-particle spin-down'', and
``right-particle spin-up, left-par\-tic\-le spin-down''.  Let us now
imagine that the pair so-created is separated sufficiently far that
classical information would take a long time to travel between
them. Then, we measure for example the $z$-component of the spin of
one of the particles (say, the left one).  This measurement has two
possible outcomes, which occur with probability one-half each,
implying that the von Neumann uncertainty of the density matrix describing
any one of the particles (denoted by subscripts $L$ and $R$),
\be
\rho_{L,R}=\frac12|\uparrow\rangle\langle\uparrow|+
           \frac12 |\downarrow\rangle\langle\downarrow| \label{eq2}
\ee
is one bit
\be
S(\rho_{L,R})=-{\rm Tr_{R,L}}\left(\rho_{L,R}\log_2\rho_{L,R}\right)=1
\ee
in spite of the fact that entropy of the combined system {\em
  vanishes}. The latter is of course well-known: for 
a quantum mechanical ``pure state'' ($\rho_{\rm
EPR}^2 = \rho_{\rm EPR}$, where $\rho_{\rm EPR} = |\Psi_{\rm
EPR}\rangle\langle\Psi_{\rm EPR}|$) the 
von-Neumann entropy {\rm vanishes} $S(\rho_{\rm
EPR})=0$, i.e., the state is perfectly well-known. 

Clearly
then, the quantum nature of the EPR state is very peculiar since the
uncertainty of a part of this system can be larger than the
uncertainty of the pair. Classically,
this is impossible. Indeed, if we describe uncertainties using
(classical) Shannon entropies, the Shannon entropy
of a system $A$, say, with $A\subset AB$, is 
\be 
H(A)\le H(AB)\;.  
\ee
This property of {\em monotonicity} of entropies is violated in
quantum mechanics~\cite{bib_wehrl}. This violation, on the other hand,
can be described consistently in an information-theoretic formalism 
which allows
for {\em negative} conditional entropies~\cite{bib_ca1,bib_ca3}. In other
words, there exists an information theory, extended to the quantum
regime, in which the violation of classical laws such as monotonicity
are inevitable consequences. 

Quantum {\em entanglement} situations, such as encountered in EPR
pairs, are prototype systems to examine the classically forbidden
regime of negative entropies.
In the case at hand, the joint, conditional, mutual, and marginal
entropies of the EPR pair can be summarized by the entropy diagram in
Fig.~1. Such diagrams are used extensively in classical information
theory and serve as mental scratch pads to remind us of the separation
of unconditional entropies into conditional and mutual pieces. While
in the past 
the violation of monotonicity prevented the use of Venn
diagrams in quantum information theory, the introduction of negative
entropies has reinstated this useful
tool~\cite{bib_ca1,bib_ca3,bib_ca2,bib_meas}. 
In particular, we can see how 
\be
S(L)\not\leq S(LR)\;.
\ee
is possible in Fig.~1 if $S(L|R)$ is negative.

\begin{figure}[t] 
\caption{ Quantum entropy diagrams. (a) Definition of
joint [$S(LR)$] (the total area), marginal [$S(L)$ or $S(R)$], 
conditional [$S(L|R)$ or $S(R|L)$] and mutual [$S(L:R)]$ entropies for a 
quantum system $LR$ separated into two subsystems $L$ and $R$; (b)  
their respective  values for the EPR pair.
\label{fig1} }
\vskip 0.25cm
\centerline{\psfig{figure=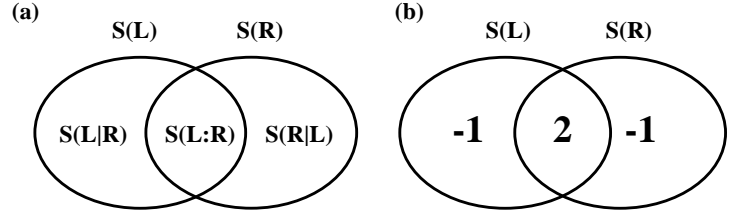,width=3.75in,angle=0}}
\end{figure}

\par
The repercussions of such an information-theoretic
description of entanglement for the extraction of information from
such a state (a measurement) are manifold. Here, we focus on
EPR experiments and other quantum paradoxes, and on implications for
physical as well as quantum pictures of reality.

\section{Information Theory of EPR Experiments}
In order to assess the relation between quantum and physical reality in
an EPR measurement, we need to describe both the quantum system (the
EPR wavefunction) {\em and} the classical devices it becomes entangled with,
using information theory.

\begin{figure}[h] 
\caption{Measurement of EPR pair $Q_1Q_2$ by devices $A_1$ and $A_2$. 
\label{meas} }
\vskip 0.25cm
\centerline{\psfig{figure=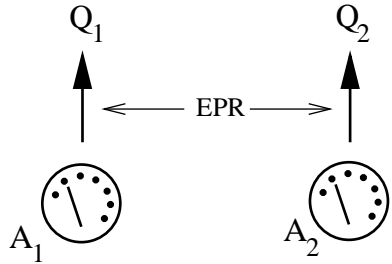,width=2.0in,angle=-90}}
\end{figure}

Let $A_1$ and $A_2$ denote measurement {\em devices}, each of the devices
measuring the $z$ component of one member of an EPR pair, for example
(see Fig.~\ref{meas}).

\begin{figure} 
\caption{Entropy diagram for the {\em devices}: (a) recording $\sigma_z$ for
each of the entangled particles, or (b) recording
$\sigma_z$ for one and $\sigma_x$ for the other particle.
\label{fig2} }
\vskip 0.25cm
\centerline{\psfig{figure=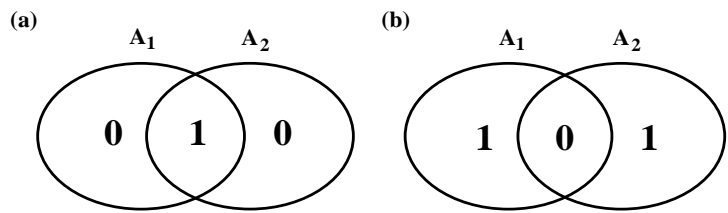,width=3.75in,angle=0}}
\end{figure}

It is an experimental fact that the measurement of the state
of one of the particles (say, $\sigma_z$) allows a 100\% accurate
prediction of what the outcome of the measurement of the other one
will be.  Thus, the outcomes of the measurement of $\sigma_z$ are
perfectly correlated, a situation described by the 
entropy diagram in Fig.~\ref{fig2}a, which appears perfectly classical (no
negative numbers appear). 

Note that the correlations between the devices are quite
unlike those of the quantum system that is measured, a peculiarity
that is quantitatively manifested when comparing Figs.~1b and
\ref{fig2}a.  The
reason why the correlations between the measurement devices (Fig.~\ref{fig2}a)
{\em incompletely} mirror the entanglement present in the quantum state
(Fig.~1b) must be due to the device's classical nature: classical
conditional entropies cannot be negative.  However, classicality 
must not be {\em assumed} for the devices, it is a mathematical result of
the information-theoretic treatment of measurement (which gives rise
to Fig.~\ref{fig2}a)~\cite{bib_meas}. 

Assume now that {\em orthogonal} spin projections are measured on the two
halves of the EPR pair, say $\sigma_z$ on the left particle, and
$\sigma_x$ on the right one. If we assume that measuring the state of
one partner confers reality to the state of the measured {\em system},
we must conclude that the experiment just described would allow us to
infer the $z$ and $x$ projections {\em simultaneously}, a state of
affairs strictly forbidden by the uncertainty relation. In their
landmark paper~\cite{bib_epr} EPR therefore
conclude that, since conventional quantum mechanics cannot describe
this peculiar situation, the theory must necessarily be incomplete.
This is the essence of the EPR paradox.  It relies on a definition of
reality based on ``certain prediction''\footnote{EPR wrote
  in~\cite{bib_epr}: ``If, without in any way disturbing a system, we
  can predict with certainty (i.e., with probability equal to unity)
  the value of a physical quantity, then there exists an element of
  physical reality corresponding to this physical quantity.''}
according to which the state of the second particle would acquire
physical reality after measuring its EPR partner.  In fact, for this
particular experiment (measuring $\sigma_z$ on the left and $\sigma_x$
on the right particle) it is found that the outcomes recorded by
the devices are completely {\em uncorrelated} as depicted by the
classical entropy diagram for the {\em devices} pictured in
Fig.~\ref{fig2}b. Rather
than reflecting an incompleteness of the formalism, these outcomes are
{\em predicted} by quantum information theory, and imply that physical
reality is attributed to the state of the second measurement {\em
  device}, or more precisely the {\em relative} state of the devices,
while there {\em cannot} be any correlation between the apparatus and the
quantum state proper (as we show below). 
In view of the importance of this conclusion,
let us repeat it once more. Quantum information theory predicts that
in EPR-type measurements, the measurement device {\em cannot} reflect
the state of of the quantum system. In the language of 
Mermin~\cite{bib_mermin}, the
correlations between the devices are real, i.e., possess physical reality,
while the quantum system itself does not.

\par Let us show this in more detail. For a proper quantum
information-theoretic analysis, we need to consider four 
systems: a quantum pair
$Q_1Q_2$ and a pair of ancillae $A_1A_2$. The ancillae can be thought of
as classical devices that are built to reflect the quantum
states. From a measurement point of view, we are interested in the
correlations between the {\em ancillae}, as only such correlations are
experimentally accessible (relative states).  
Before we analyze them using {\em quantum}
entropy diagrams, let us ponder what we expect to find from an
orthodox point of view. 

\begin{figure} 
\caption{{\em Classical} entropy diagram for the EPR measurement of 
spin-projections: (a)  both devices measure $\sigma_z$, (b) one device
measures $\sigma_z$, the other $\sigma_x$. Note that the entropy of
$A_1$ and $A_2$ have to be one bit in each case, as the measurement
outcomes are equiprobable, while $Q_1Q_2$ is thought to have {\em two}
independent equiprobable degrees of freedom.  
\label{fig3} }
\vskip 0.25cm
\centerline{\psfig{figure=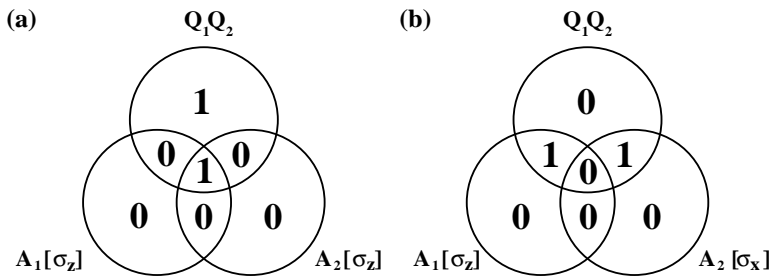,width=4.0in,angle=-90}}
\end{figure}

One of the fundamental tenets of classical
measurement theory is that a measurement device is constructed such as
to mirror the state of the object to be measured as accurately as
possible. In other words, measurement entails the transfer of this
information to a macroscopic system that is more suited to accurate
observation, without altering the state of the system. While it is
well-known that {\em quantum} measurements cannot be made without
altering the quantum state~\cite{note_noncloning}, the general belief
is that the quantum state {\em after} measurement {\em is} truthfully 
portrayed by the
device. In other words, it is believed that correlations between
the quantum state and the ancillae in the measurement situation allow
the extraction of information about the quantum system.  Let us
consider the ``orthodox'' (classical) entropy diagram
(Fig.~\ref{fig3}) for an EPR
measurement, drawing the quantum system $Q_1Q_2$ as one system,
measured by the ancillae $A_1$ and $A_2$.
These diagrams reveal the paradox inherent in this
description.  On the one hand, when the same projection of the spin
(e.g., $\sigma_z$) is measured for {\em both} particles (Fig.~\ref{fig3}a)
classical reasoning suggests that the quantum system and the
measurement devices {\it share} information (one bit in the center of
the diagram). On the other hand, when orthogonal polarizations are
measured (Fig.~\ref{fig3}b) the measurement devices must appear
uncorrelated. According to a ``physical realism'' or ``hidden
variable'' picture, both diagrams in Fig.~\ref{fig3} must have a common
underlying classical diagram relating five ensembles: the EPR pair
$Q_1Q_2$ and the four possible measurements $A_1[\sigma_z]$,
$A_1[\sigma_x]$, $A_2[\sigma_z]$, and
$A_2[\sigma_x]$\footnote{The diagrams in Fig.~\ref{fig3} are obtained from such an
underlying diagram by ignoring two out of the five variables: Fig.~\ref{fig3}a
by ignoring $A_1[\sigma_x]$ and $A_2[\sigma_x]$, Fig.~\ref{fig3}b by ignoring 
$A_1[\sigma_x]$ and $A_2[\sigma_z]$. ``Ignoring'' a system is achieved
by the mathematical operation of tracing it out of the 
joint density matrix. }.
 This underlying diagram, however,
is in contradiction with the Heisenberg uncertainty principle, as it
implies that the {\em counterfactual} variables $\sigma_x$ and
$\sigma_z$ (of the same particle) can be measured together.  Thus,
this classical treatment of information leads to a paradox.

\par 
\begin{figure}[t] 
\caption{ (a) {\em Quantum} entropy diagram for the EPR measurement of same 
spin-projections: e.g., $A_1$ and $A_2$ both measure $\sigma_z$. (b) 
reduced diagram obtained by tracing over the quantum states $Q_1$ and $Q_2$.
\label{fig4} }
\vskip 0.25cm
\centerline{\psfig{figure=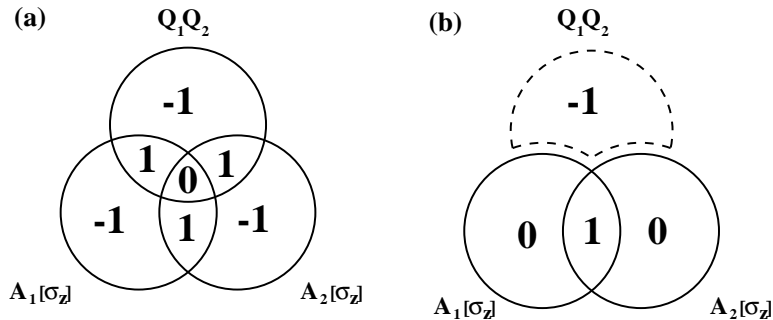,width=4.0in,angle=-90}}
\end{figure}
\begin{figure} 
\caption{ (a) Quantum entropy diagram for the EPR measurement of orthogonal 
spin-projections, e.g., $A_1$ measures $\sigma_x$ while $A_2$ records 
$\sigma_z$. (b) Reduced diagram as above.
\label{fig5} }
\vskip 0.25cm
\centerline{\psfig{figure=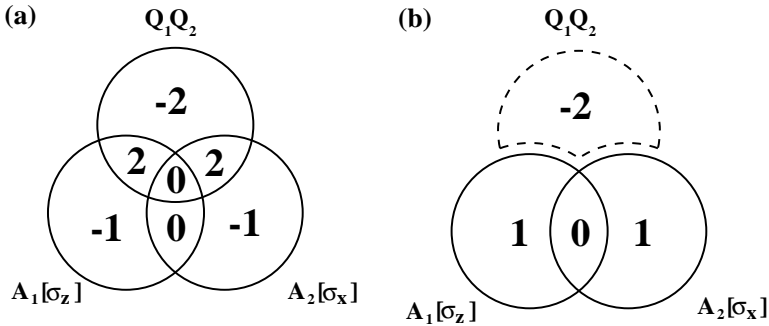,width=4.0in,angle=-90}}
\vskip -0.5cm
\end{figure}

The paradox is resolved by drawing the {\em quantum} entropy diagrams
for the measurements (Figs.~\ref{fig4} and \ref{fig5}). 
The values for the respective
quantum entropies entering these diagrams can be obtained by
straightforward calculation~\cite{bib_meas}. In Fig.~\ref{fig4} the entropy
diagram representing the situation where the same polarizations are
measured is that of a GHZ state~\cite{bib_ghz}: fully symmetric and
maximum quantum
entanglement between three entities.  As is well-known, tracing over
(ignoring) one member of such a triplet produces classical
correlations (of the type depicted in Fig.~\ref{fig2}a) in the remaining
doublet, as indicated in Fig.~\ref{fig4}b. As a consequence, the quantum
entropy diagram of Fig.~\ref{fig4}a {\em correctly} reproduces the observed
correlations between the detectors $A_1$ and $A_2$.  Closer inspection
of Fig.~\ref{fig4}a, however, reveals that while the measurement devices are
perfectly correlated as they should, their mutual entropy (the single
bit of
information gained in the measurement) is {\em not} shared by the
quantum system $Q_1Q_2$.  In Figs.~\ref{fig4} and \ref{fig5}, this
ternary mutual
information\footnote{Just like any entropy, {\em information}, which
  is the mutual entropy between {\em two} systems, can be split up
  into a conditional and a mutual piece with respect to a third
  system~\cite{bib_shannon}.}
is represented by the center of the diagram, and measures
how much of the correlations between the measurement devices is shared
by the quantum system. If the center of the diagram is zero, we must
conclude that no
information is shared between quantum system and classical devices.

The same is true for the measurement situation in Fig.~\ref{fig5}a, where
{\em incompatible} polarizations are measured. Again, the (four part)
system is fully
entangled, and ignoring the quantum state produces the experimentally
observed independence of the measurement results (Fig.~\ref{fig5}b, compare
Fig.~\ref{fig2}b).  Yet, the correlation between quantum state and measurement
device (the mutual information between the measuring and the measured
system) is {\em unchanged} from the previous arrangement, in fact, it
vanishes in both cases\footnote{The mutual information between quantum
  system and both classical devices also vanishes for any intermediate
  situation between Figs.~\ref{fig4} and \ref{fig5}, since the joint
  system $Q_1Q_2A_1A_2$ is always a pure state~\cite{bib_meas}.}. 

This situation leads us to
suggest that we must abandon at least one cherished notion of orthodox
measurement theory: that the apparatus necessarily reflects the state
of the system it was built to measure, by being {\em correlated} with it
in the sense of Shannon. Rather, it is the correlations {\em between}
the ancillae
(the reality of their {\em relative} state)
that create the {\em illusion} of measurement.  Indeed, any subsequent
measurement on each side (left or right), for example,  
would yield the {\em same} result, over
and over again, while still not implying {\em anything} about the
quantum wavefunction. Each observer that repeats a measurement becomes
classically correlated to the earlier outcome, {\em whatever the outcome}.
Still, the quantum reality of the superposition is unperturbed by
these measurements: none of the repeatable measurements yield any
information about the quantum state, while they are internally
completely consistent. Note that the orthodox interpretation of these
correlations involves the concept of a wavefunction collapse: the
measurement of the first particle projects---or collapses---the
wavefunction of the other one, to account for the perfect
correlation. Since the devices do not reflect the state of the quantum
system, however, no collapse is needed to explain the correlations,
nor does it actually occur, as we now show.

\section{Information Theory of Schr\"odinger Cats}
The Schr\"odinger-cat paradox is of prime importance for the
understanding of quantum decoherence and the quantum-classical
boun\-da\-ry. The latter have received increased attention recently due to
their importance for the design of quantum computation and
communication devices~\cite{bib_qcomp}.
\par
The Schr\"odinger-cat paradox explores the relationship between classical
and quantum variables by coupling them together in such a way that the
decay of a radioactive substance (say, one isolated atom) implies the
demise of a cat locked up with the deadly contraption in a sealed
room. The quantum
reality of the (isolated) atomic system is that of a superposition of
a decayed atom with gamma ray, and an undecayed atom without. If
brought into contact with the cat, however, quantum mechanics forces
us to include the cat in this entangled wavefunction
\begin{equation}
|\Psi\rangle =\frac1{\sqrt2}\left(|A^\star,0,L\rangle +
|A,1,D\rangle\right)\;,
\end{equation}
where $|A^\star\rangle$ and $|0\rangle$ refer to the excited atom and
{\em absent} gamma, while $|A\rangle$ and $|1\rangle$ are the
wavefunctions of the decayed atom and the gamma.  Furthermore,
$|L\rangle$ and $|D\rangle$ refer to the ``live'' and ``dead'' cat
eigenstates.  The paradox arises if an observer peeks into the room to
observe the state of the cat. Does the cat's wavefunction immediately
collapse into one of its eigenstates (dead or alive) upon observation?
The preceding analysis teaches us that this is not necessary.  The
observer can be thought of as a fourth system that is now {\em quantum
entangled} with the previous troika: atom, gamma, and cat
\begin{equation}
|\Psi\rangle =\frac1{\sqrt2}\left(|A^\star,0,L,l\rangle +
|A,1,D,d\rangle\right)\;,
\end{equation}
introducing ``observer eigenstates'' $|l\rangle$ and $|d\rangle$.  
Then, upon tracing over the quantum degrees of freedom of the atom (after all,
this experiment involves monitoring the cat and not the atom), the cat
(serving as the hapless gamma-ray detector) appears perfectly
correlated with the observer peeking into the room. Cat and observer
agree, so to speak, about the observation, and their state is entirely
classical. Yet, their agreement is completely decorrelated, {\em
disjoint}, from the
quantum state, as their {\em mutual} information shared with the atomic
system vanishes. In other words, the classical reality displayed by
cat and observer does not imply anything about the quantum reality of
atom and gamma ray, or vice versa. Fundamentally, the reason why the
observer does not register a cat mired in a quantum superposition of
the living and non-living states is because the observer, having
interacted with the cat, is entangled with, and thus part of, the
{\em same} wavefunction. As the wavefunction is {\em indivisible}, an
observer (or measurement device) would have to monitor {\em itself} in 
order to learn about the wavefunction. This is logically impossible.

\section{Conclusions}
To summarize, we assert that quantum reality is ``real'' in the
sense that quantum mechanics completely and deterministically
describes the evolution of a closed system (not just its
wavefunction), and that the statistical character arises from the fact
that an observer, because he is part of the closed system, is offered
an {\em incomplete} view of the quantum system he attempts to
measure. Consequently, the quantum universe is deterministic as
Einstein's physical reality demands, but must include the observer as
one of its parts due to the inseparability of entangled quantum
states. The recent information-theoretic analysis of quantum
measurement~\cite{bib_ca2,bib_meas} shows that such an observer indeed
perceives the system he is measuring as probabilistic, and thus that
Bohr's complementarity principle emphasizing the importance of the
system/observer relation therefore holds at the same time.  If quantum
reality is so elusive, how then can we learn about its nature?
Fortunately, while negative entropy cannot be reflected in classical
instruments directly, it is possible to infer it from combined
measurements and comparison with classical bounds (a case in point are
Bell inequalities~\cite{bib_bell}, see also~\cite{bib_entbell}). 
Thus, quantum reality does leave
its traces in experiments, while the direct observation of superpositions
is impossible.
\vskip 0.5cm 
\noindent{\bf \large Acknowledgments}
\vskip 0.25cm

\noindent This work was supported in part by NSF Grant Nos. PHY 94-12818 and PHY
94-20470 and by a grant from DARPA/ARO through the QUIC program
(\#DAAH04-96-1-3086).  N.J.C.\  is Collaborateur Scientifique of the
Belgian National Fund for Scientific Research. An earlier version of
this paper was circulated in the Fall of 1996 under the title
``Physical Reality and Quantum Paradoxes''.


\begin{thebibliography}{99}
\bibitem{bib_epr} A. Einstein, B. Podolsky, and N. Rosen, 
Can quantum-mechanical description of physical reality be considered complete? 
Phys. Rev. {\bf 47}, 777 (1935).

\bibitem{bib_bohrnat} N. Bohr, The quantum postulate and the recent
  development of atomic theory,  Nature {\bf 121}, 580 (1928).

\bibitem{bib_wheeler} J.A. Wheeler and W.H. Zurek, eds., 
{\it Quantum Theory and Measurement} (Princeton University Press, 1983).

\bibitem{bib_bellbook} J.S. Bell, {\it Speakable and Unspeakable in
Quantum Mechanics} (Cambridge University Press, Cambridge, 1987).

\bibitem{bib_schommers} W. Schommers, ed., {\it Quantum Theory and
Pictures of Reality}, (Springer, Berlin, 1989).

\bibitem{bib_mittel} P. Busch, P.J. Lahti, and P. Mittelst\"adt, 
{\it The Quantum Theory of Measurement} (Springer, New York,
1991). 

\bibitem{bib_cushing} J.T. Cushing, {\it Quantum Mechanics---Historical
Contingencies and the Copenhagen Hegemony}, (University of Chicago
Press, Chicago, 1994).

\bibitem{bib_bohm2} D. Bohm, A suggested interpretation of the
quantum theory in terms of hidden variables, I and II,
Phys. Rev. {\bf 85}, 166 (1952). 

\bibitem{bib_bell} J.S. Bell, On the Einstein Podolsky Rosen paradox,
 Physics {\bf 1}, 195 (1965).

\bibitem{bib_vn}J. von Neumann, {\it Mathematische Grundlagen der
Quantenmechanik} (Springer Verlag, Berlin, 1932).


\bibitem{bib_schroed} E. Schr\"odinger, Die gegenw\"artige Situation
  in der Quantenmechanik, Naturwissenschaften {\bf 23}, 807
(1935). 

\bibitem{bib_monroe} C. Monroe, D. M. Meekhof, B.E. King, and D. J. Wineland,
A Schr\"odinger cat superposition state of an atom, 
Science {\bf 272}, 1131 (1996); J. J. Slosser and G.J. Milburn,
Creating metastable Schr\"odinger cat states,
Phys. Rev. Lett. {\bf 75}, 418 (1995).
 
\bibitem{bib_shannon} C.E. Shannon and W. Weaver, {\it The Mathematical 
Theory of Communication} (University of Illinois Press, 1949). 

\bibitem{bib_bohr} N. Bohr, Can quantum-mechanical description of physical 
reality be considered complete?, 
Phys. Rev. {\bf 48}, 696 (1935).

\bibitem{bib_ca2} N.J. Cerf and C. Adami, Quantum mechanics of
  measurement, eprint quant-ph/9605002, unpublished.

\bibitem{bib_meas} N.J. Cerf and C. Adami, Information theory of
  quantum entanglement and measurement, Physica D (1998).  

\bibitem{bib_mermin} N.D. Mermin, What is quantum mechanics trying
  to tell us?, eprint quant-ph/9801057. 

\bibitem{bib_bohm} D. Bohm, {\it Quantum Theory}, (Prentice-Hall, Englewood 
Cliffs, 1951), pp. 611-623.

\bibitem{bib_wehrl}A. Wehrl, General properties of entropy, 
Rev. Mod. Phys. {\bf 50}, 221 (1978).

\bibitem{bib_ca1} N.J. Cerf and C. Adami, Negative entropy and
  information in quantum mechanics,  Phys. Rev. Lett. {\bf 79} 
(1997) 5194.

\bibitem{bib_ca3} N.J. Cerf and C. Adami, Negative entropy in quantum
    information theory, in {\it New Developments on
    Fundamental Problems in Quantum Physics}, Fundamental Theories of
    Physics {\bf 81}, M. Ferrero and
    A. van der Merwe, eds. (Kluwer Academic Publishers, Dordrecht, 1997) p. 77.


\bibitem{note_noncloning} This is the
essence of the quantum non-cloning theorem, see W.K. Wootters and W.H. Zurek, 
A single quantum cannot be cloned, 
 Nature {\bf 299}, 802 (1982); D. Dieks, 
Communication by EPR devices, Phys. Lett. {\bf 92A}, 271 (1982).

\bibitem{bib_ghz} D.M. Greenberger, M.A. Horne, and A. Zeilinger, 
Going beyond Bell's theorem, in {\it Bell's Theorem, Quantum Theory, 
and Conceptions of the Universe}, M. Kafatos, ed., (Kluwer, Dordrecht,
1989) p. 69; N.D. Mermin, Quantum mysteries revisited, 
Am. J. Phys. {\bf 58}, 731 (1990).

\bibitem{bib_qcomp} D.P. DiVincenzo, Quantum computation, 
Science {\bf 270}, 255 (1995);
I.L. Chuang, R. Laflamme, P.W. Shor, and W.H. Zurek, Quantum
computers, factoring, and decoherence, {\it ibid.},
p. 1633; C.H.Bennett, Quantum information and computation, 
Phys. Today {\bf 48}, 24 (October, 1995). 

\bibitem{bib_entbell}N.J. Cerf and C. Adami, Entropic Bell
    inequalities, Phys. Rev. {\bf A 55}, 3371 (1997). 

\end{thebibliography}
\end{document}